\documentclass[twocolumn,showpacs,preprintnumbers,amsmath,amssymb]{revtex4}

\usepackage[dvips]{graphicx}
\usepackage{dcolumn}
\usepackage{bm}
\usepackage{color}

\begin{document}

\preprint{APS/123-QED}

\title{Rumor Evolution in Social Networks}

\author{Yichao Zhang$^{1,2}$}

\author{Shi Zhou$^{2}$}
\email{s.zhou@cs.ucl.ac.uk}

\author{Zhongzhi Zhang$^{3,4}$}
\email{zhangzz@fudan.edu.cn}

\author{Jihong Guan$^{1}$}
\email{jhguan@tongji.edu.cn}

\author{Shuigeng Zhou$^{3,4}$}
\email{sgzhou@fudan.edu.cn}


\affiliation{$^{1}$Department of Computer Science and Technology,
Tongji University, 4800 Cao'an Road, Shanghai 201804, China}

\affiliation{$^{2}$Department of Computer Science, University
College London, Gower Street, London, WC1E 6BT, United Kingdom}

\affiliation {$^{3}$School of Computer Science, Fudan University,
Shanghai 200433, China}

\affiliation {$^{4}$Shanghai Key Lab of Intelligent Information
Processing, Fudan University, Shanghai 200433, China}

\date{\today}

\begin{abstract}
Social network is a main tunnel of rumor spreading. Previous studies are concentrated on a static rumor spreading. The content of the rumor is invariable during the whole spreading process. Indeed, the rumor evolves constantly in its spreading process, which grows shorter, more concise, more easily grasped and told. In an early psychological experiment, researchers found about $70\%$ of details in a rumor were lost in the first $6$ mouth-to-mouth transmissions~\cite{TPR}. Based on the facts, we investigate rumor spreading on social networks, where the content of the rumor is modified by the individuals with a certain probability. In the scenario, they have two choices, to forward or to modify. As a forwarder, an individual disseminates the rumor directly to its neighbors. As a modifier, conversely, an individual revises the rumor before spreading it out.
When the rumor spreads on the social networks, for instance, scale-free networks and small-world networks, the majority of individuals actually are infected by the multi-revised version of the rumor, if the modifiers dominate the networks.
Our observation indicates that the original rumor may lose its influence in the spreading process. Similarly, a true information may turn to be a rumor as well. Our result suggests the rumor evolution should not be a negligible question, which may provide a better understanding of the generation and destruction of a rumor.
\end{abstract}


\maketitle

\section{INTRODUCTION\protect}

Rumor spreading is a fundamental topic in psychology~\cite{TPR} and sociology~\cite{LC169}. In the past decade, rumor spreading on social networks, e.g., small-world networks~\cite{PRE64050901,PRE65041908,PRE69066130}, and scale-free networks~\cite{PHYSICAA374457,PRE76036117,PRE69066130,PRE69055101},
has attracted lots of attention from physical and sociological communities~\cite{NATURE2041118,PRE76036117,EPL7868005,ARXIV0505031,PRL103038702,PRE78066110}.
In the small world network, a threshold $P_c$ of spreading in the rewiring probability of links was reported. As the probability is larger than $P_c$,
i.e., when the average path length of the network is short enough, rumors can be disseminated globally. On the other hand, in the scale-free networks,
the discussions are mainly focused on the spreading efficiency and the final infected ratio, which was analytical solved by a mean-field approximation~\cite{PHYSICAA374457}. Most recently, a model describing two propagating rumors with different probabilities of acceptance is also worth a mention~\cite{PRE81056102}.

Simultaneously, the development of the dynamical model itself is also pretty heuristic~\cite{PRL92218701,PRE73046138}. Some researchers took consideration of the accumulation effect of rumor in the process of persuading the ignorant~\cite{PRL92218701} and others investigated the degeneration of information on a spatial system~\cite{PRE73046138}. Both of the works provide valuable insight into the rumor dynamics.

In real social networks, the rumor spreading process is much more complicated than the reported scenarios~\cite{BMB15523,JMS8265,SN2755,PRE64046134,PRE65036107}. The rumor evolves constantly in its spreading process, which grows shorter, more concise, more easily grasped and told~\cite{TPR}. The behavior originates from the cumulative modifications during the spreading process, which is called 'Chinese Whispers' or 'Telephone' in some conditions~\cite{TMM}. The phenomenon is easy to be observed in real social networks. For example, on a microblogging site Twitter, once a user discusses a certain topic, his or her followers will understand the topic indirectly. If it is a rumor, the following discussions can roughly be classified as: affirmative, negative, curious, unrelated and unknown arguments~\cite{SMA}. No matter which class the following tweets belong to, they are still the variants of the original rumor. Thus, whether the original rumor can infect the whole networks depends not only on the existence of connections among individuals but also on their strategies. We divide the strategies into two classes, to forward it directly or to change it before spreading it out. For a convenience, the individuals forwarding rumor are denoted by forwarders. Conversely, the individuals changing rumor are denoted by modifiers. In real email systems, modifiers can be not only users but also machines, called remailer~\cite{JTLP6175}.

In this paper, we will investigate rumor spreading in social networks, where individuals have two static behaviors to forward and to modify. As long as receiving a rumor, a forwarder will deliver it to its neighbors directly, while an ignorant modifier will deliver the rumor after revising once. When spreading in a network including modifiers, the rumor probably has to experience a series of revisions. Even if the original rumor is an unconfirmed truth, the spreading process may turn it into a rumor as well. Thus, this general model may draw a relatively complete picture of real rumor spreading, including generation and destruction of a rumor. Previous works taking all the individuals as forwarders~\cite{PRE64050901,PRE65041908,PRE69066130,PHYSICAA374457,PRE76036117,PRE69055101} is a special case of our law. In the previous models~\cite{PRE64050901,PRE65041908,PRE69066130,PHYSICAA374457,PRE76036117,PRE69055101}, the spreading rate and the annihilation rate are two key parameters, which governs the spreading process and the final infected ratio. In this paper, they are not what we concern. Thus, we set both of them 1 to focus on the evolution of the rumor itself. Given that the evolution of rumor content is a sociological topic, we only focus on the revised frequency here.
\begin{figure}
\scalebox{0.5}[0.5]{\includegraphics[trim=60 0 0 0]{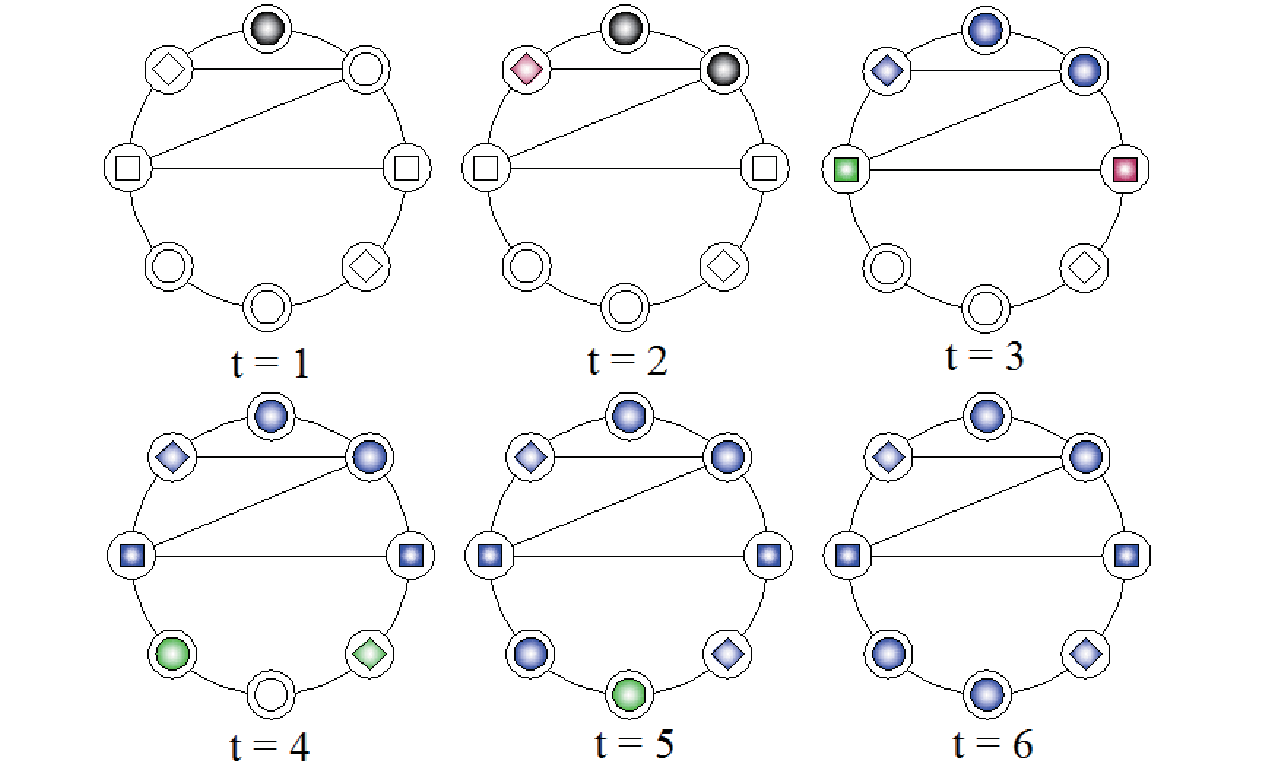}}
\caption{(Color online) Illustration of rumor spreading in a network with forwarders and modifiers.
Circles denote the forwarders and squares denote the modifiers.
We set a rumor starts with black, whose color changes once when encountering a modifier.
Black denotes the original rumor, red denotes the rumor modified once, and green denotes the rumor modified twice.
Blue denotes stifler. The illustration represents the whole process of rumor spreading in six time steps.}
\label{smile}
\end{figure}

\section{Rumor Evolution on Social Networks}

As discussed in the original rumor model (the DK model)~\cite{NATURE2041118}, individuals can play three roles: ignorants, spreaders, and stiflers, whose densities are denoted by $i(t)$, $s_{x}(t)$, and $r(t)$, respectively. Here, $s_{x}$ denotes the version $x$ $(x=1,2,...,n)$ of a rumor. The original version is the version $1$. We set the normalization condition $i(t)+s_{sum}(t)+r(t)=1$ and $s_{sum}=\Sigma_{x} s_{x}$.
The role of an individual start with an ignorant. If this individual is infected before the rumor vanishes, it will turn to be a spreader. Finally, it will be a stifler once it spreads the rumor to a spreader or stifler. If this individual is not infected during the process, it will keep its role. Besides these dynamical roles, each individual also has two types of static behaviors, to forward or to modify. The behaviors of individuals are fixed in the spreading process. We will discuss all the possible cases in what follows.

When receiving a certain version of the rumor, an ignorant forwarder becomes a spreader of the version. When receiving two or more different versions of the rumor, the forwarder accepts the latest version. For example, given $a>b$, if the forwarder receives the version $a$ and $b$ at the same step, its role becomes $S_{a}$.

Once receiving the rumor, an ignorant modifier revises it before disseminating it out. When receiving two or more different versions of the rumor at the same step, the modifier revises the latest version. For example, given $a>b$, if the modifier receives the version $a$ and $b$ at the same step. Its role becomes $S_{a+1}$.

If a spreader receives the original rumor or the revised versions in the following steps, it will turn to be a stifler, no matter whether it is a forwarder or modifier. It is simply because the spreader has disseminated a similar information to its neighbors, there is no need to do that again~\cite{PRE69066130}. On the other hand, the neighbors who send the versions of the rumor to it, would be not interested in the similar information as well. They will turn to be stiflers at the same step. To explain this point clearly, a simple illustration is shown in Fig.~\ref{smile}.

\subsection{Rumor Evolution on Scale-free Networks}
When a rumor is injected into the heterogeneous networks as BA networks~\cite{SCI286509}, the set of coupled properties can be written as what follows. Consider an ignorant forwarder $i$ with degree (or connectivity) $k$ after $t$ steps.
When receiving the rumor $x$, a probability with which it becomes a spreader $s_x$ is
\begin{equation}
P_{i\rightarrow s_{x}}^{is}(t,k)=F
kP(k)i_k(t)\Sigma_{k'}\frac{k'P(k')s_{k'}(t)}{\langle k\rangle},
\end{equation}
where $P(k)$ denotes the degree distribution of the networks. $F$ denotes the forwarders' fraction. If $i$ is an ignorant modifier, a probability with which it becomes a revised rumor spreader $s_{x+1}$ is
\begin{equation}
P_{i\rightarrow s_{x+1}}^{is}(t,k)=(1-F)
kP(k)i_k(t)\Sigma_{k'}\frac{k'P(k')s_{k'}(t)}{\langle k\rangle}.
\end{equation}
A probability with which a spreader $s_x$ becomes a stifler $r$ is
\begin{equation}
P_{s_{x}\rightarrow r}^{sr}(t,k)=kP(k)s_k(t)\Sigma_{k'}\frac{k'P(k')\left[s_{k'}(t)+r_{k'}(t)\right]}{\langle{k}\rangle}.
\end{equation}

We define
\begin{equation}
 \langle{R_{k}}\rangle=\frac{\sum_{i\in\{i|degree(i)=k\}}i_R}{N_k},
\end{equation}
where $i_R$ denotes the last version of the rumor at individual $i$ before individual $i$ turns to be a stifler, where $i=1,2,...,N$ and $R=1,2,...$.
$\langle R_{k} \rangle$ represents the frequency that a rumor has been revised on average before annihilating at an individual with degree $k$. The rate equation for the average revised frequency $\langle R_{k} \rangle$ on degree $k$ can be written as:
\begin{equation}
\frac{d\langle{R_k(t)}\rangle}{dt}=(1-F)P(k)ki_k(t)\Sigma_{k'}\frac{k'P(k')s_{k'}(t)\langle{R_{k'}(t)}\rangle}{\langle{k}\rangle}.
\end{equation}
The evolution of the densities $s_k(t)$ and $r_k(t)$ satisfy
the following set of coupled differential equations:
\begin{equation}
\frac{di_k(t)}{dt}=-kP(k)i_k(t)\Sigma_{k'}\frac{k'P(k')s_{k'}(t)}{\langle
k\rangle}.
\end{equation}
\begin{eqnarray}
\frac{ds_k(t)}{dt}=&kP(k)i_k(t)\Sigma_{k'}\frac{k'P(k')s_{k'}(t)}{\langle
k\rangle}&\nonumber\\
&-kP(k)s_k(t)\Sigma_{k'}\frac{k'P(k')\left[s_{k'}(t)+r_{k'}(t)\right]}{\langle
k\rangle}&.
\end{eqnarray}
\begin{equation}
\frac{dr_k(t)}{dt}=kP(k)s_k(t)\Sigma_{k'}\frac{k'P(k')\left[s_{k'}(t)+r_{k'}(t)\right]}{\langle
k\rangle}.
\end{equation}

\subsection{Rumor Evolution on Small-world Networks}
When a rumor is injected into the homogeneous networks as WS networks~\cite{NATURE393440}, the set of coupled properties can be written as what follows. Consider an ignorant forwarder $i$ after $t$ steps. When receiving the rumor $x$, a probability with which it becomes a spreader $s_x$ is
\begin{equation}
P_{i\rightarrow s_{x}}^{is}(t)=F\langle k\rangle i(t)s(t).
\end{equation}
If $i$ is an ignorant modifier, a probability with which it becomes a revised rumor spreader $s_{x+1}$ is
\begin{equation}
P_{i\rightarrow s_{x+1}}^{is}(t)=(1-F)\langle
k\rangle i(t)s(t).
\end{equation}
A probability with which a spreader $s_x$ becomes a stifler $r$ is
\begin{equation}
P_{s_{x}\rightarrow r}^{sr}(t)=\langle k\rangle
s(t)\left[s(t)+r(t)\right].
\end{equation}

Thus, in this case, the rate equation for the average revision frequency $R(t)$ can be written as:
\begin{equation}
\frac{d\langle{R(t)}\rangle}{dt}=(1-F)\langle{k}\rangle~i(t)s(t)\langle{R(t)}\rangle,\label{R}
\end{equation}
where the evolution of the densities $s(t)$ and $r(t)$ satisfy
the following set of coupled differential equations:
\begin{equation}
\frac{di(t)}{dt}=-\langle k\rangle i(t)s(t).\label{i}
\end{equation}
\begin{equation}
\frac{ds(t)}{dt}=\langle k\rangle s(t)\left[i(t)-\left(s(t)+r(t)\right)\right].\label{s}
\end{equation}
\begin{equation}
\frac{dr(t)}{dt}=\langle k\rangle s(t)\left[s(t)+r(t)\right].\label{r}
\end{equation}
In the infinite time limit, previous studies~\cite{PHYSICAA374457,PRE69066130} show that the fraction of individuals infected by the rumor can be written as
\begin{equation}
r(\infty)=1-e^{-2r(\infty)}.\label{r_whisper}
\end{equation}
In this scenario, the ratio is a constant, which is only determined by Eq.~\ref{i},~\ref{r} and the normalization condition $i(\infty)+r(\infty)=1$ for $s_{sum}(\infty)=0$. More to the point, one can find that the final infected ratio $r(\infty)$ is irrelevant to the evolution of the rumor. For the evolution, the system of differential equations Eq.~\ref{R} and~\ref{i} can be analytically solved. With $\langle{R(0)}\rangle=(\frac{F}{N}+\frac{2(1-F)}{N})$, one can derive
\begin{equation}
\langle{R(t)}\rangle=\exp((F-1)i(t)+(1-F)\frac{N-1}{N}-\ln{\frac{2-F}{N}}).
\end{equation}
In the infinite time limit, we have
\begin{equation}
\langle{R(\infty)}\rangle=\exp((F-1)(1-r(\infty))+(1-F)\frac{N-1}{N}-\ln{\frac{2-F}{N}}).
\end{equation}
Considering $r(\infty)$ is a constant, $\langle{R(\infty)}\rangle$ only depends on the forwarders' fraction $F$ and the total number of individuals $N$ for the WS networks.

\section{Distributions of Revised Frequencies}
\begin{figure*}
\scalebox{0.3}[0.3]{\includegraphics[trim=0 0 50
0]{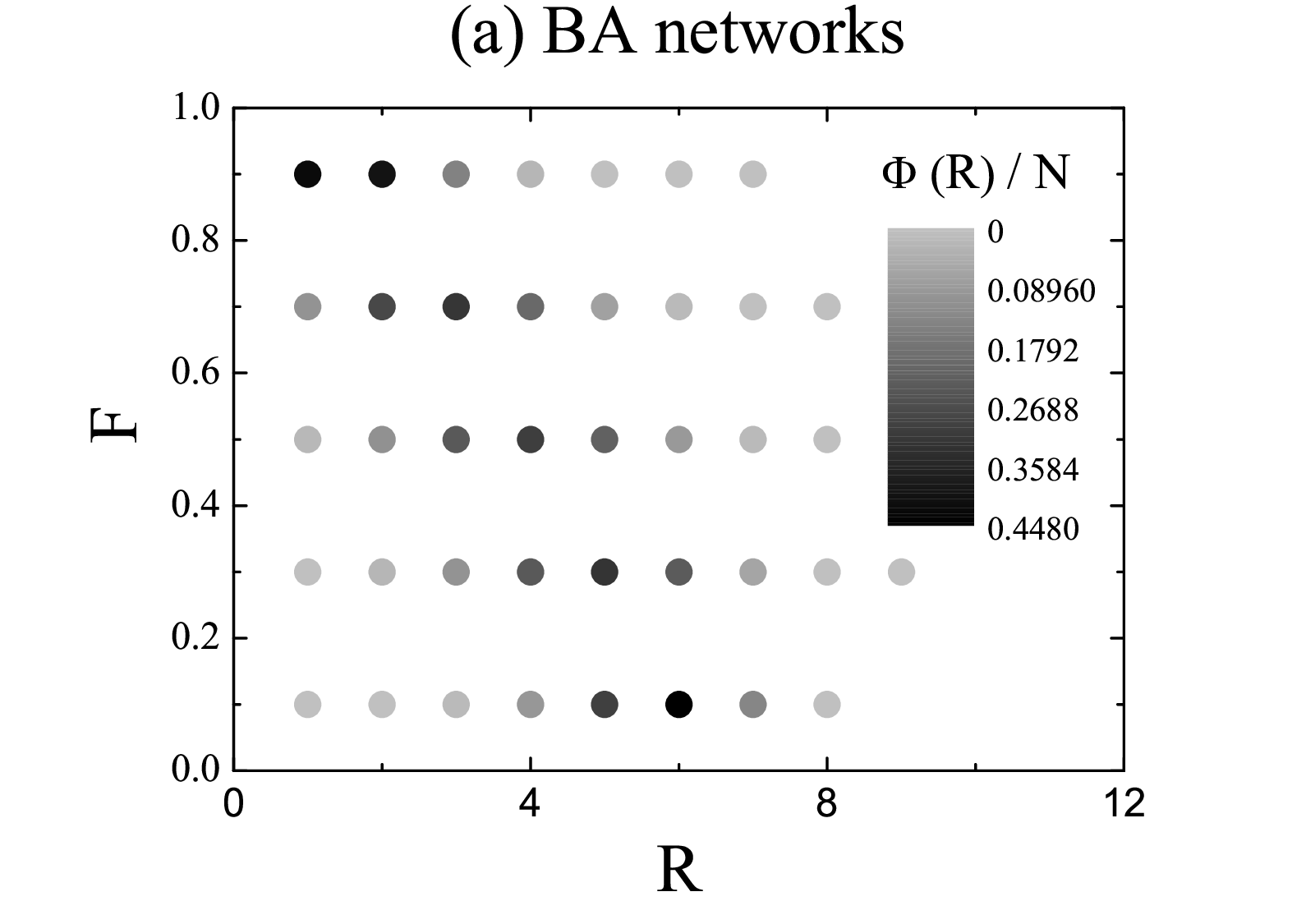}}
\scalebox{0.3}[0.3]{\includegraphics[trim=50 0 0
0]{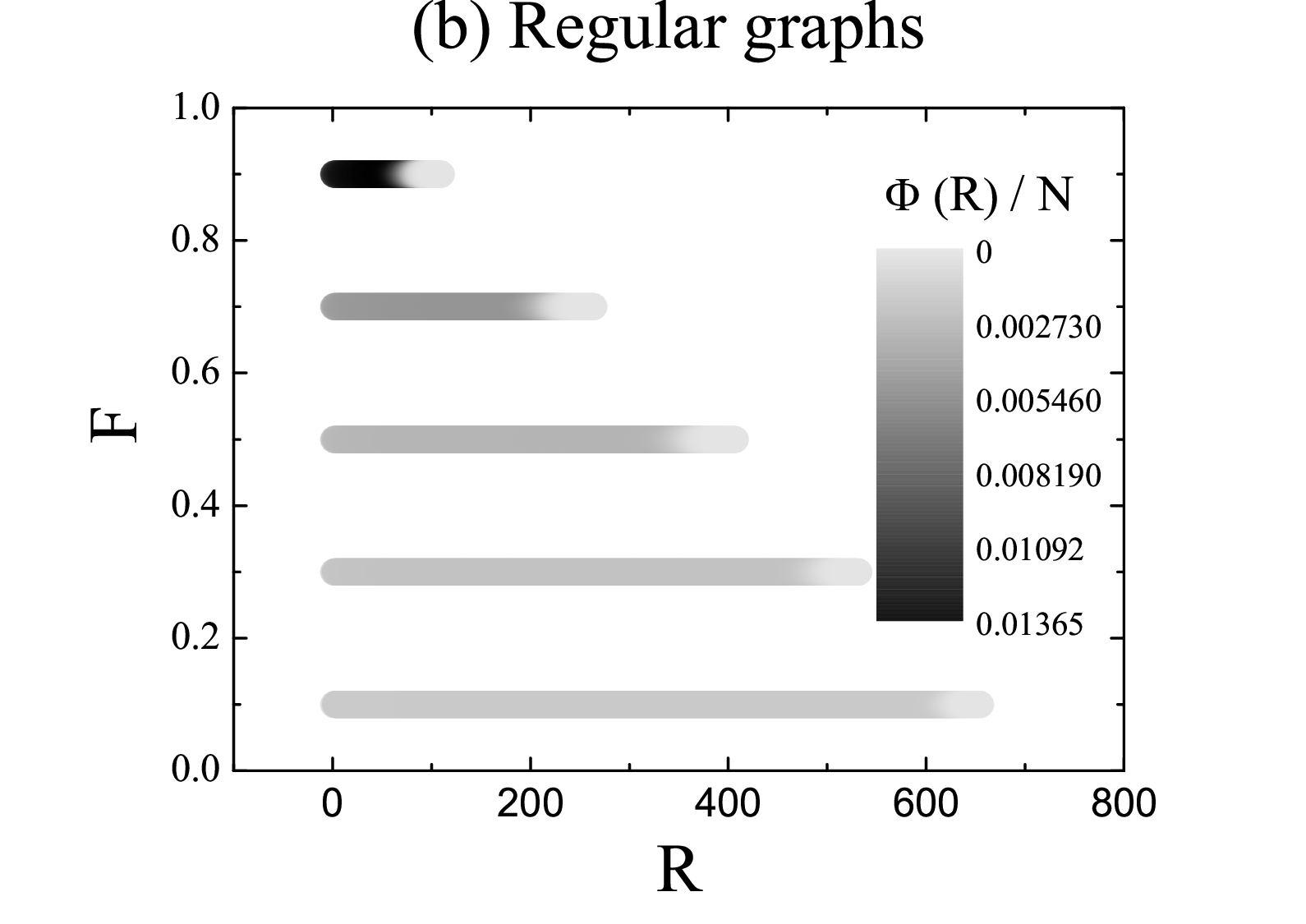}}
\scalebox{0.3}[0.3]{\includegraphics[trim=0 0 50
0]{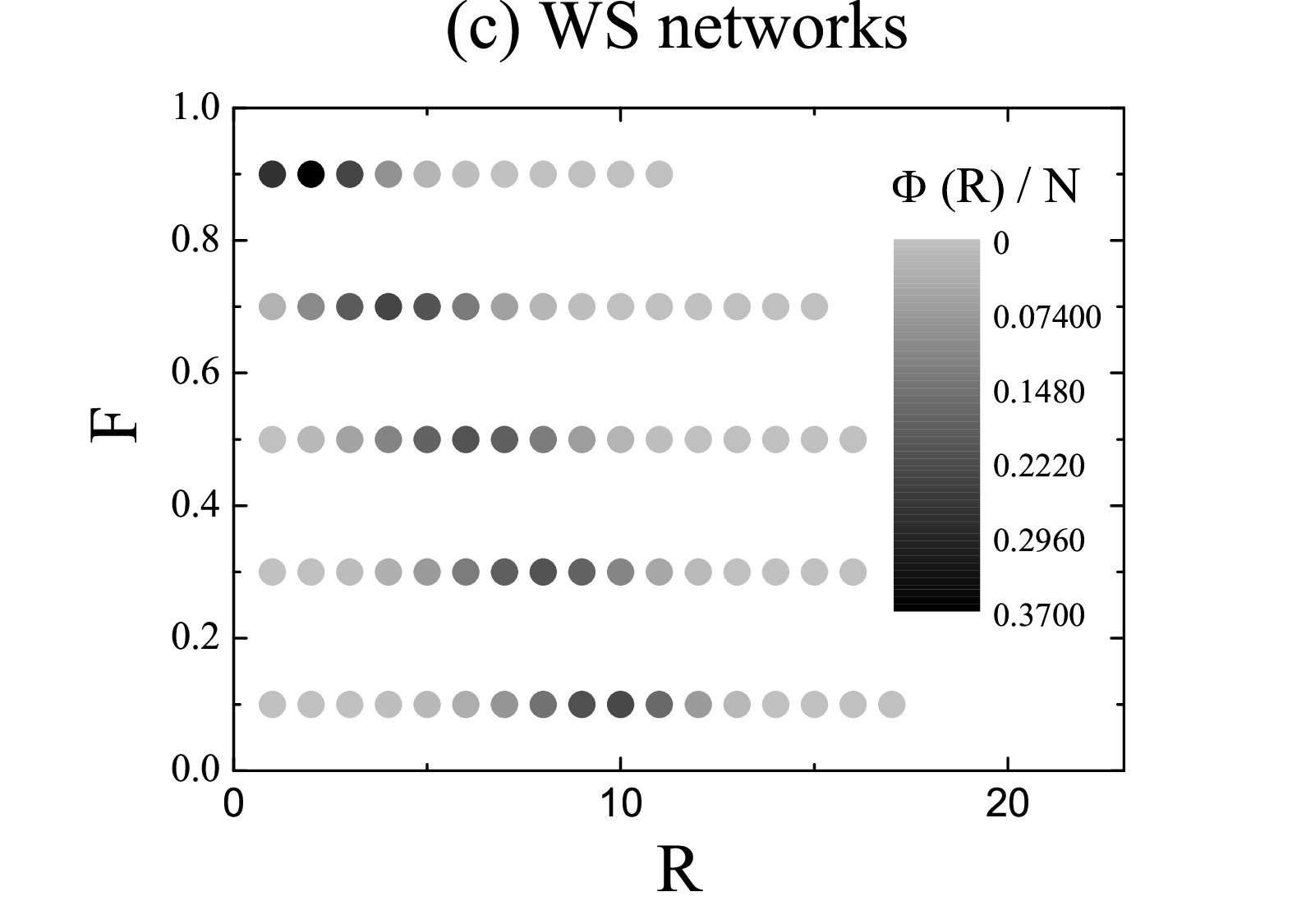}}
\scalebox{0.3}[0.3]{\includegraphics[trim=50 0 0
0]{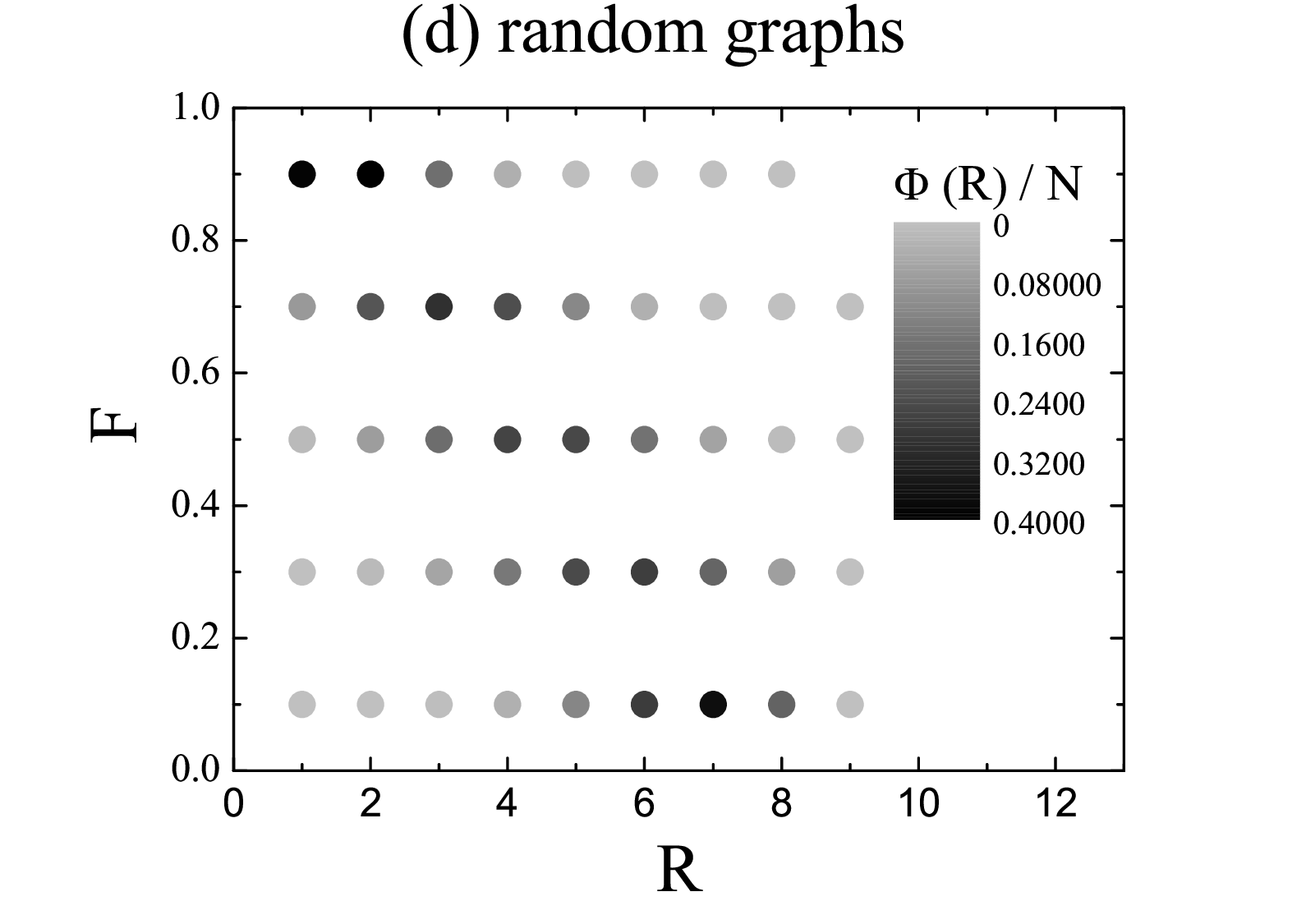}}
 \caption{$\Phi(R)/N$ as a function of $R$ and $F$ for different topologies with $F=0.1,0.3,0.5,0.7,0.9$. $N$ denotes the size of the networks, which is $4096$ in our simulations. ({\bf a}) shows the simulation results obtained on BA scale-free networks, which generated by $m_0=m=3$~\cite{SCI286509}; ({\bf b}) shows the simulation results obtained on regular graphs, which are formed by 4,096 identical individuals of degree 6; ({\bf c}) shows the simulation results obtained on WS small-world networks, which are generated by randomly rewiring 10\% of the links in the regular graphs; ({\bf d}) shows the simulation results obtained on random graphs, which are generated by randomly rewiring all the links in the regular graphs~\cite{NATURE393440}. With the initial conditions $i(0)=\frac{N-1}{N}$, $s(0)=\frac{1}{N}$, and $r(0)=0$, simulation results were obtained by ten random assignments of modifiers on ten different realizations of the same type of network specified by the appropriate parameters. Each plot in this figure corresponds to $100*N_k(k=6)$ simulations. For each run, we set a randomly picked individual with degree $6$ as the first spreader.}~\label{CW_revision_dis}
\end{figure*}

To clarify the result of evolution, we then run extensive simulations on both the WS and BA networks for five different values of $F$. Also, we investigate the effects of totally regular and random structures. We generate ten WS networks, BA networks, regular graphs (networks), and random graphs (networks) using random seeds. In Fig.~\ref{CW_revision_dis}, we measure the distributions of $R$. We define $\Phi(R)$ as the number of individuals who were the spreaders of the rumor revised $R$ times before the rumor vanishes.

As shown in Fig.~\ref{CW_revision_dis}(a)(c)(d), for the BA networks, one can find out the position of the maximum of the distribution of $R$ shifts to the left with $F$. Simultaneously, all the maximums grow with $F$.
In Fig.~\ref{CW_revision_dis}(b), for the regular graphs, one can observe that the distribution is relatively uniform. The positions of the maximum shift to the left with $F$ as well. Unlikely, the versions of the rumor are much richer. One can observe that the maximum grows with $F$, but the growth is relatively limited.

For $F$ is close to $0$, the distributions of $R$ on all the networks tend to reach a relatively uniform status. In Fig.~\ref{CW_revision_dis}(b) one can observe that the number of individuals infected by various versions are basically identical for regular graphs. In Fig.~\ref{CW_revision_dis}(a)(c)(d), one can observe that the majority of the individuals are infected by the versions revised more than $4$ times.
For $F$ is close to $1$, Fig.~\ref{CW_revision_dis}(a)(b)(c)(d) show that the majority of the individuals are infected by the versions revised less than $3$ times. This observation indicates that the original rumor can keep its influence on the individuals only when most of them are forwarders in the social networks.

\begin{figure*}
\scalebox{0.3}[0.3]{\includegraphics[trim=0 10 50
0]{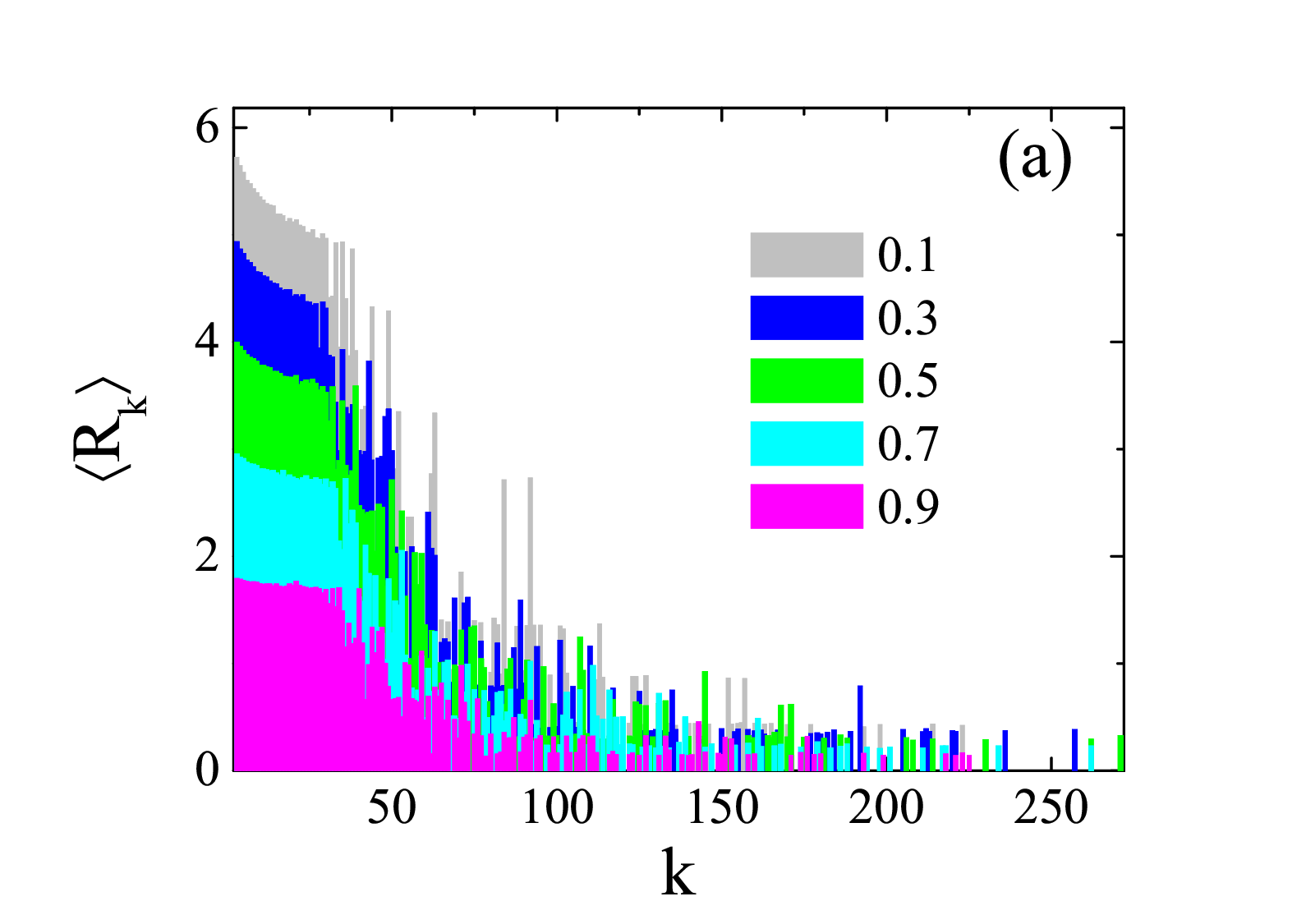}}
\scalebox{0.3}[0.3]{\includegraphics[trim=50 0 0
0]{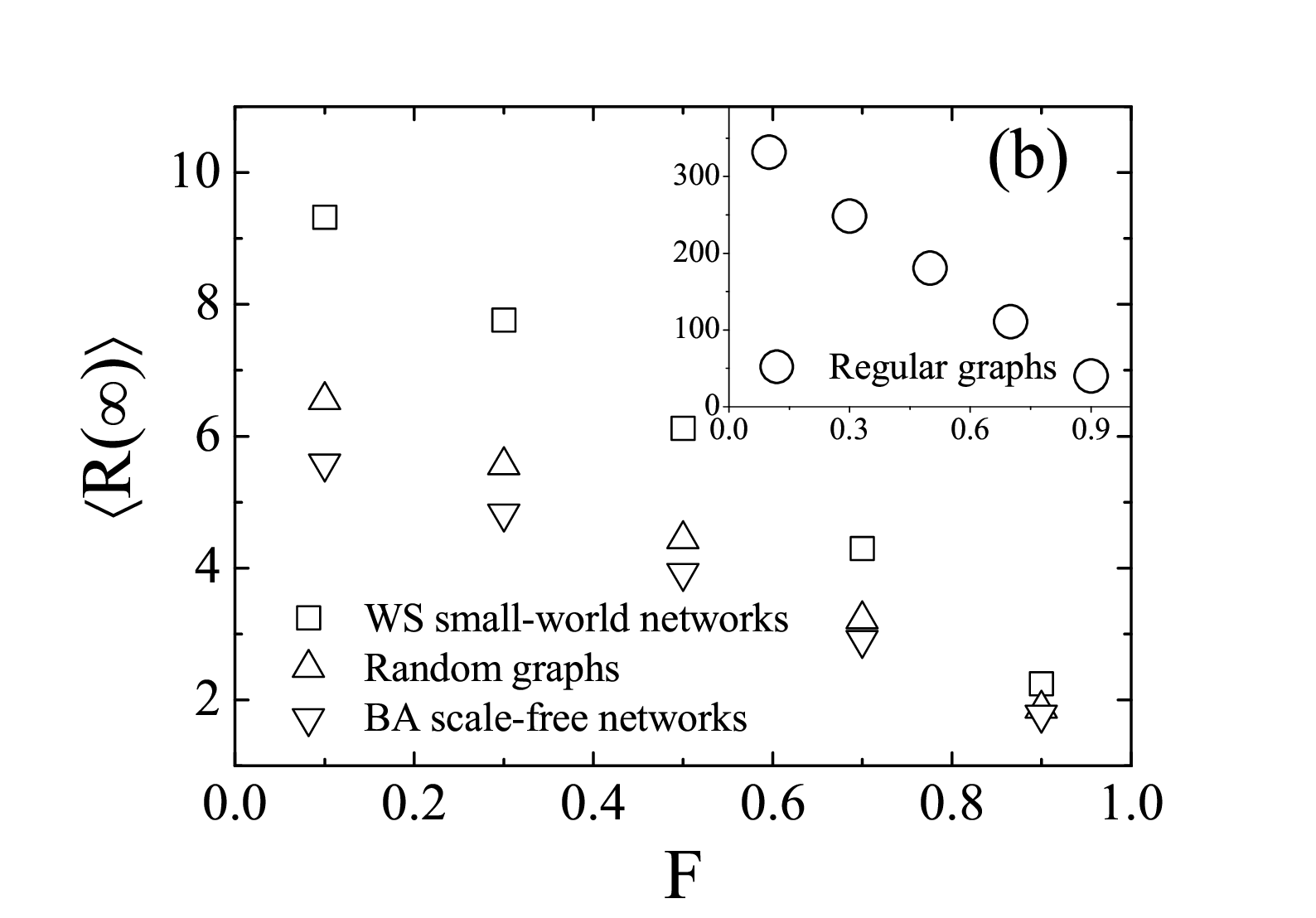}}
\scalebox{0.3}[0.3]{\includegraphics[trim=0 0 50
0]{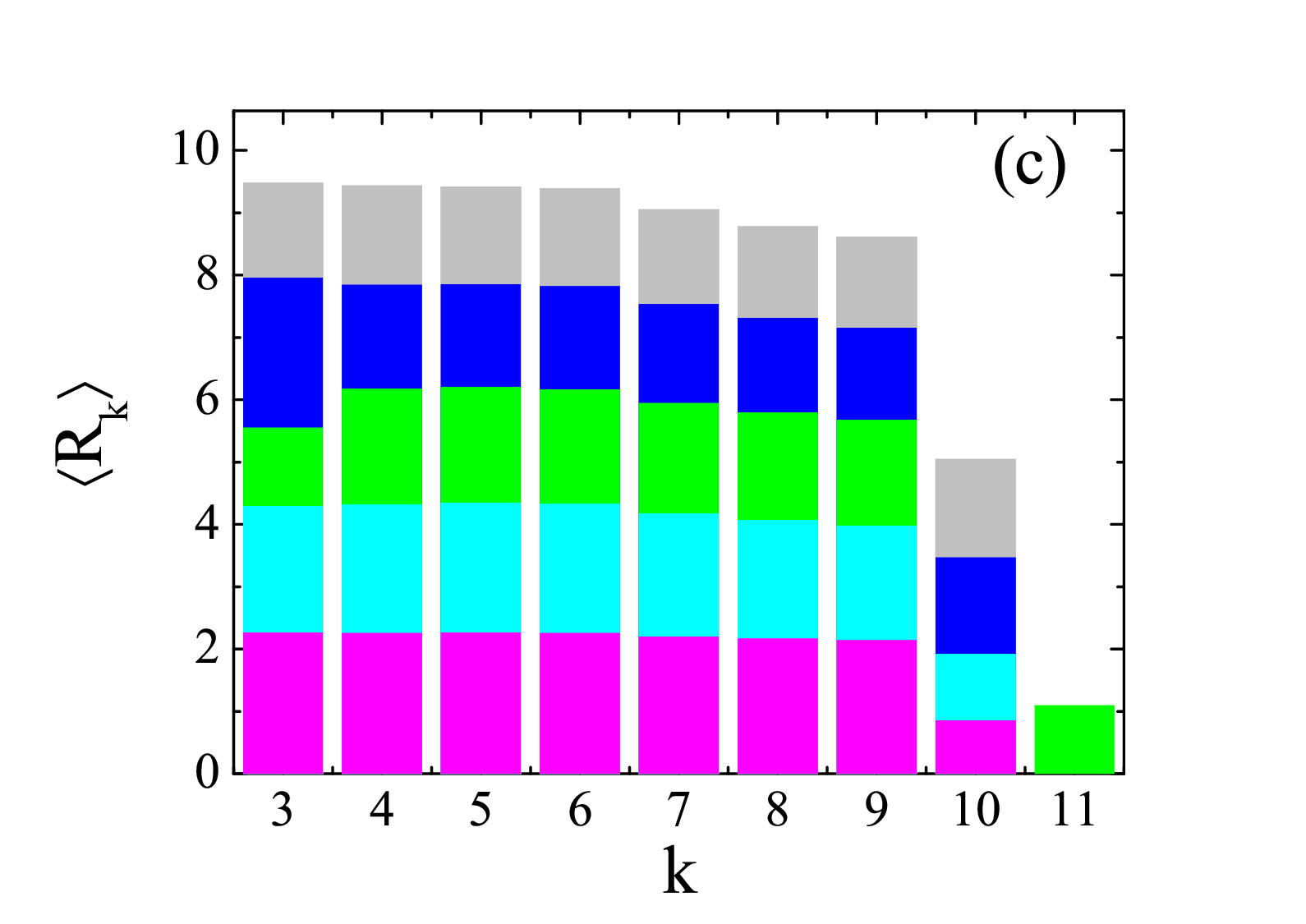}}
\scalebox{0.3}[0.3]{\includegraphics[trim=50 0 0
0]{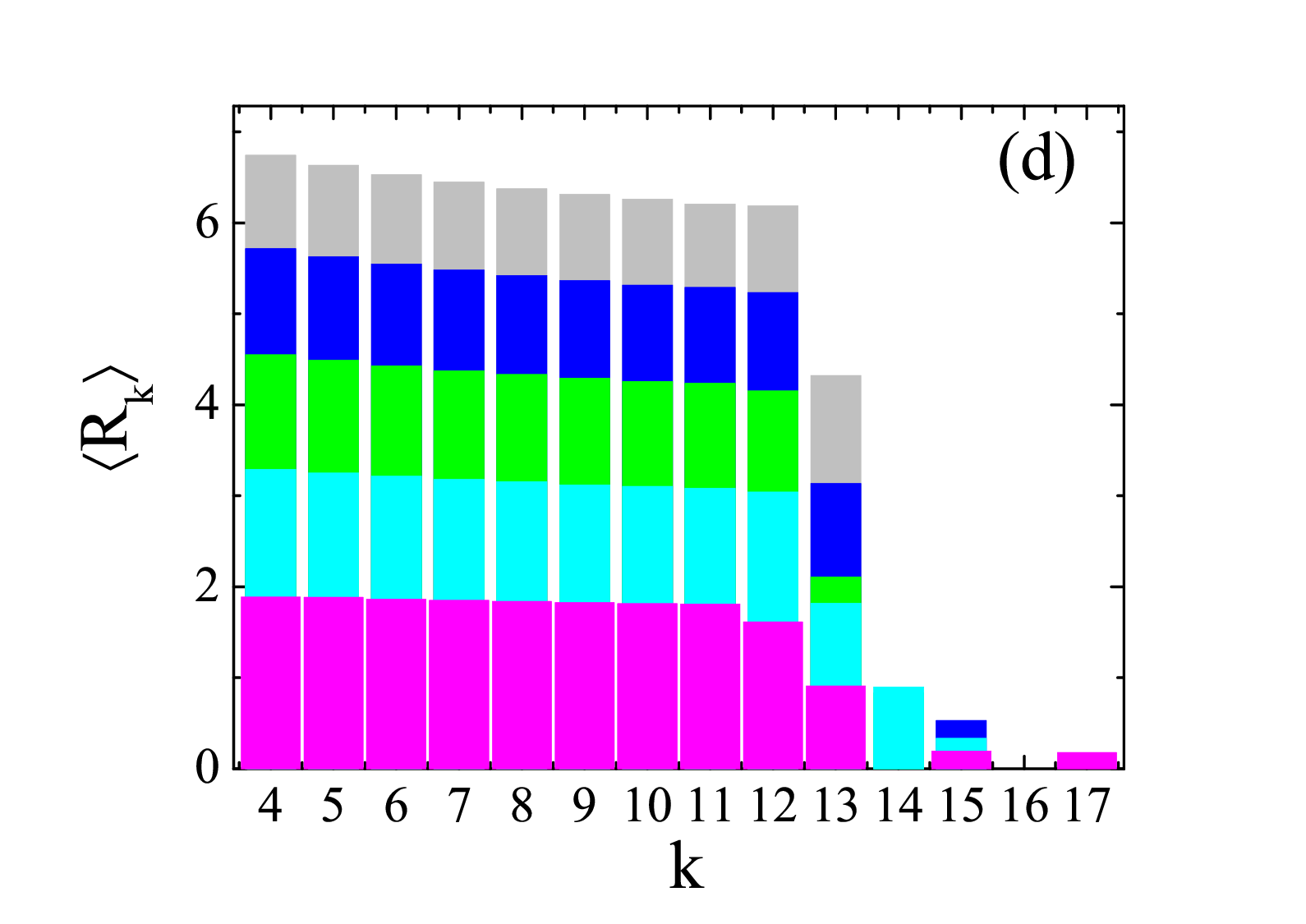}}
\caption{(Color online) $\langle R_k\rangle$ as a function of individuals' degree $k$ for different topologies with $F = 0.1, 0.3, 0.5, 0.7, 0.9$. ({\bf a}), ({\bf c}), and ({\bf d}) show simulation results for BA scale-free networks, WS small-world networks, and random graphs. ({\bf b}) shows the relations between $\langle R(\infty)\rangle$ and $F$ for regular graphs, WS networks, random graphs, and BA networks; Panel ({\bf c}) and ({\bf d}) share the same legend of $F$ with Panel ({\bf a}).}~\label{CW_revision_dis_degree}
\end{figure*}

To clarify the relation between the revised frequency and the topological feature of networks, we measure the average revised frequency $\langle R_{k} \rangle$ for the individuals with degree $k$. As shown in Fig.~\ref{CW_revision_dis_degree}(a), one can observe the individuals with low degree have a relatively high $\langle R_k\rangle$ in the BA networks. The feature decays with $F$.
For the regular graphs, all the individuals' degree is $6$. Hence, we only show the relation between $\langle R(\infty)\rangle$ and $F$ in Fig.~\ref{CW_revision_dis_degree}(b). In Fig.~\ref{CW_revision_dis_degree}(b), one can observe that $\langle R\rangle$ decays with $F$ for all the topologies investigated in this paper. The inset shows the relation between $\langle R_6(\infty)\rangle$ and $F$ for the regular graphs. Obviously, The speed of decay for the regular graphs is much higher than that of the others.
In Fig.~\ref{CW_revision_dis_degree}(c)(d), for the WS networks and random graphs, one can observe the distribution of $\langle R_k\rangle$ is close to a uniform distribution for the individuals with small degree.
For the BA scale-free networks, $\langle R_k\rangle$ is directly proportional to $N_k$. As a function of $k$, $\langle R_k\rangle$ follows a power-law. Interestingly, for the WS small-world networks and random graphs, the distribution of $\langle R_k\rangle$ is irrelevant to the degree distribution generally.

\section{CONCLUSION\protect}
To sum up, rumor as a common social phenomenon has been investigated on various topological models. In previous studies, the content of a rumor is set to be invariable in its spreading process. Indeed, most rumors evolve constantly, which may grow shorter, more concise, more easily grasped and told. It may also be commented or questioned. In this paper, we have proposed a rumor model on social networks, where two static behaviors of individuals, to modify or to forward, govern the evolution of the rumor. As defined in the previous models, each individual may have three dynamical roles, ignorant, spreader and stifler. Initially, we inject a rumor into a network, where all the individuals are ignorant. When an ignorant forwarder receives a rumor, it becomes a spreader and spreads the rumor to its neighbors directly. When an ignorant modifier receives a rumor, it becomes a spreader as well. However, it revises the content before spreading. If it receives the rumor or the revised version again in the following steps, it will turn to be a stifler. This is because it disseminated a similar information to its neighbors before, the neighbors may lose interest in the information. When all the individuals are forwarders, our model can be reduced to the previous rumor models.

We have run extensive simulations to investigate the distributions of the revised frequency $R$ on various topological structures.
For the BA scale-free networks, regular graphs, WS small-world networks and random graphs, we found that the position of the maximum of the distributions shifts to the left with the fraction of forwarders $F$. For a small $F$, the distributions on all the networks tend to reach a relatively uniform status. The majority of the individuals are infected by the multi-revised versions. For a large $F$, the original rumor can keep its influence on the individuals.

To clarify the relation between the revised frequency and topological structure, we have measured the average revised frequency $\langle R_{k} \rangle$ as a function of degree $k$.
For the BA scale-free networks, $\langle R_k\rangle$ is directly proportional to the number of individuals with degree $k$. Instead, for the WS small-world networks and random graphs, $\langle R_k\rangle$ doesn't depend on degree distribution generally.
For the regular graphs, respecting all the individuals have an identical degree, we have measured the relation between the final average revised frequency for all the individuals $\langle R(\infty)\rangle$ and $F$. We found $\langle R(\infty)\rangle$ decays dramatically with $F$ in this regular structure. The speed of decay for regular graphs is much higher than that in the other structures.

As a common social phenomenon, rumor evolution has been highly accelerated by modern information networks. The evolving rumor model we propose in this paper can provide a more realistic framework for the future research of rumor dynamics. We believe our results may provide a better understanding of rumor spreading in real social networks. Our observations are also capable of promoting related studies on the agent-based rumor spreading.\smallskip

\begin{acknowledgments}
This research was supported by the National Natural Science Foundation of China under Grant Nos. 860873040 and 60873070, and 863 program under Grant No. 2009AA01Z135. Jihong Guan was also supported by the ``Shuguang Scholar" Program of Shanghai Education Development Foundation under grant No. 09SG23.
\end{acknowledgments}

\bibliography{REGSN}

\end{document}